\documentclass{mn2e}
\input epsf

\usepackage{graphicx}
\usepackage{color}

\renewcommand{\vec}[1]{ {\bmath #1} }

\title[Width of cold fronts in clusters of galaxies]{On the
width of cold fronts in clusters of galaxies due to conduction}

\author[Xiang, Churazov, Dolag, Springel, Vikhlinin]{F.~Xiang$^{1}$,
E.~Churazov$^{1,2}$, K.~Dolag$^{1}$, V.~Springel$^{1}$, A.~Vikhlinin$^{3}$ \\
$^1$ Max-Planck-Institut f\"ur Astrophysik, Karl-Schwarzschild-Strasse 1, 85741
Garching, Germany\\
$^2$ Space Research Institute (IKI), Profsoyuznaya 84/32, Moscow 117810,
Russia\\
$^3$ Harvard-Smithsonian Center for Astrophysics, 60 Garden St.,
Cambridge, MA 02138 \\
}

\pagerange{\pageref{firstpage}--\pageref{lastpage}}
\pubyear{2001}

\begin{document}
\maketitle

\label{firstpage}
\begin{abstract}
We consider the impact of thermal conduction in clusters of galaxies
on the (unmagnetized) interface between a cold gaseous cloud and a
hotter gas flowing over the cloud (the so-called cold front). We argue
that near the stagnation point of the flow conduction creates a
spatially extended layer of constant thickness $\Delta$, where
$\Delta$ is of order $\sim\sqrt{kR/U}$, and $R$ is the curvature
radius of the cloud, $U$ is the velocity of the flow at infinity, and
$k$ is the conductivity of the gas. For typical parameters of the
observed fronts, one finds $\Delta \ll R$. The formation time of such
a layer is $\sim R/U$. Once the layer is formed, its thickness only
slowly varies with time and the quasi-steady layer may persist for
many characteristic time scales. Based on these simple arguments one
can use the observed width of the cold fronts in galaxy clusters to
constrain the effective thermal conductivity of the intra-cluster
medium.
\end{abstract}


%

\sloppypar

\section{Introduction}
Chandra observations of galaxy clusters often show sharp
discontinuities in the surface brightness of the hot intra-cluster
medium (ICM) emission (Markevitch et al., 2000, Vikhlinin, Markevitch,
Murray, 2001, see Markevitch \& Vikhlinin 2007 for a review). Most of
these structures have lower temperature gas on the brighter (higher
density) side of the discontinuity, contrary to the expectation for
non-radiative shocks in the ICM. Within the measurement uncertainties,
the pressure is continuous across these structures, suggesting that
they are contact discontinuities rather than shocks.  In the
literature these structures are now called ``cold fronts''.

There are several plausible mechanisms responsible for the
formation of such cold fronts, all of them involving relative
motion of the cold and hot gases. Below we will consider the case
of a hot gas flow over a colder gravitationally bound gas cloud,
which is a prototypical model of a cold front. In such a situation
one expects that ram pressure of the hotter gas strips the outer
layers of the colder cloud, exposing denser gas layers and forming
a cold front near the stagnation point of the hot flow (Markevitch
et al., 2000, Vikhlinin et al., 2001a, Bialek, Evrard and Mohr,
2002, Nagai \& Kravtsov, 2003, Acreman et al., 2003, Heinz et al.,
2003, Asai, Fukuda \& Matsumoto, 2004, 2007, Mathis et al., 2005,
Tittley \& Henriksen, 2005, Takizawa, 2005, Ascasibar \&
Markevitch 2006) .

Some of the observed cold fronts are remarkably thin. For example, the
width of the front in Abell 3667 (Vikhlinin et al., 2001a) is less
than $5\,{\rm kpc}$, which is comparable to the electron mean free
path. Given that the temperature changes across the front by a factor
of $\sim 2$, thermal conduction (if not suppressed) should strongly
affect the structure of the front (e.g. Ettori \& Fabian, 2000). In
fact, suppression of conduction by magnetic fields is likely to happen
along the cold front since gas motions on both sides of the interface
may produce preferentially tangential magnetic field, effectively
shutting down the heat flux across the front (e.g. Vikhlinin et al.,
2001b, Narayan \& Medvedev, 2001, Asai et al., 2004, 2005, 2007,
Lyutikov 2006). While magnetic fields are hence likely to play an
important role in shaping cold fronts, it is still interesting to
consider the expected structure of a cold front in the idealized case
of an unmagnetized plasma.

The structure of this paper is as follows. In Section~2, basic
equations are listed and a toy model of a thermally broadened
interface between cool and hot gas is discussed.  In Section~3, we
present the results of numerical simulations of hot gas flowing past a
cooler gas cloud.  In Section~4, we discuss how limits on the
effective conductivity can be obtained for the observed cold
fronts. Finally, we summarize our findings in Section~5.

\section{Thermal conduction near the stagnation point of the flow}

\subsection{Basic equations}

We parameterize the isotropic thermal conductivity $k$ as
\begin{equation}
k=f\times k_{0},
\label{eq:k}
\end{equation}
where $f<1$ is the suppression coefficient of the conductivity
relative to the conductivity $k_{0}$ of an unmagnetized plasma
(Spitzer 1962, Braginskii 1965):
\begin{equation}
k_{0}=4.6\times10^{13} \left(\frac{T}{10^8\, {\rm K}}\right)^{5/2}
\left(\frac{\rm{ln}\Lambda}{40}\right)^{-1} {\rm
erg~cm^{-1}~s^{-1}~K^{-1} }, \label{eq:k0}
\end{equation}
where $T$ is the gas temperature, and $\rm{ln}\Lambda$ is the Coulomb
logarithm.

If the scale length of temperature gradients is much larger than the
particle mean free path, then saturation of the heat flux (Cowie \&
McKee, 1977) can be neglected and the evolution of the temperature
distribution can be obtained by solving the mass, momentum and energy
conservation equations with the heat diffusion term $\vec{\nabla} \cdot k
\vec{\nabla} T$ in the energy equation (e.g. Landau \& Lifshitz, 1959):
\begin{eqnarray}
&&\frac{\partial \rho}{\partial t}+\vec{\nabla} \cdot (\rho {\vec v})
=0\label{eq:hydro1},
\\
&&\frac{\partial {\vec v}}{\partial t}+\left( {\vec v} \cdot \vec{\nabla}
\right){\vec v} =-\frac{1}{\rho}\vec{\nabla} p + g,\\
&&\frac{\partial}{\partial t} \left ( \frac{\rho {\vec v}^2}{2}
+\rho\epsilon \right )= \vec{\nabla} \cdot k \vec{\nabla} T - \vec{\nabla} \cdot \rho
{\rm v} \left ( \frac{{\vec v}^2}{2}+\omega\right ),
\label{eq:hydro}
\end{eqnarray}
where $\rho$ is the gas density, $p$ is the gas pressure, $g$ is the
gravitational acceleration, and $v$ is
the gas velocity. We adopt an ideal gas with $\gamma=5/3$, where
$\epsilon=\frac{1}{\gamma-1}\frac{k_{B}T}{\mu m_p}$,
$\omega=\frac{\gamma}{\gamma-1}\frac{k_{B}T}{\mu m_p}$, and
$p=\frac{\rho}{\mu m_p}k_BT$.

In the next section we first consider the simplified case of passive
scalar diffusion in a time independent velocity flow, while in
Section~\ref{sec:num} we discuss numerical solutions of the above
equations.

\subsection{Toy model}
\label{sec:toy}
\begin{figure*}
\includegraphics[width=0.66\columnwidth]{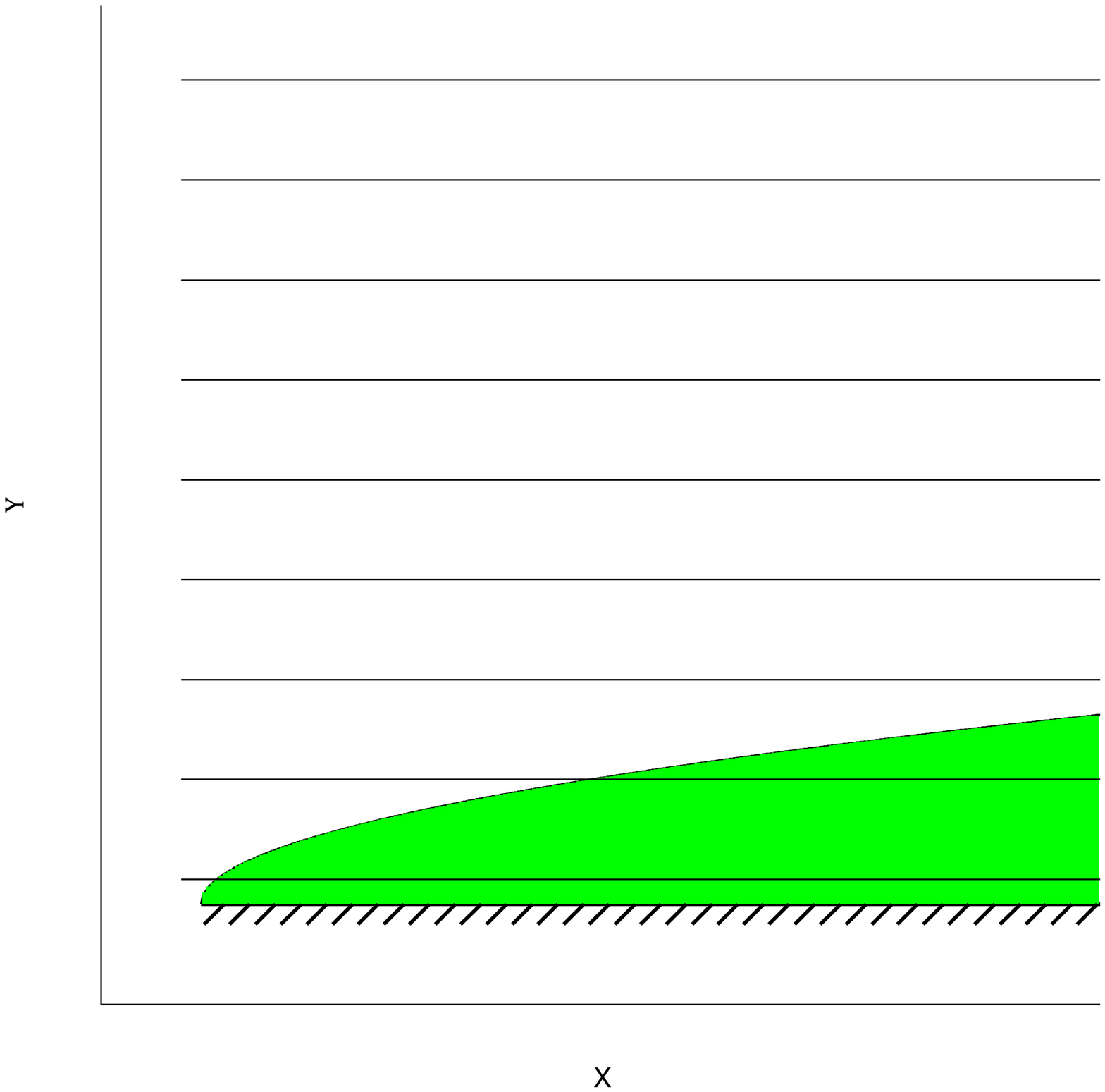}
\includegraphics[width=0.66\columnwidth]{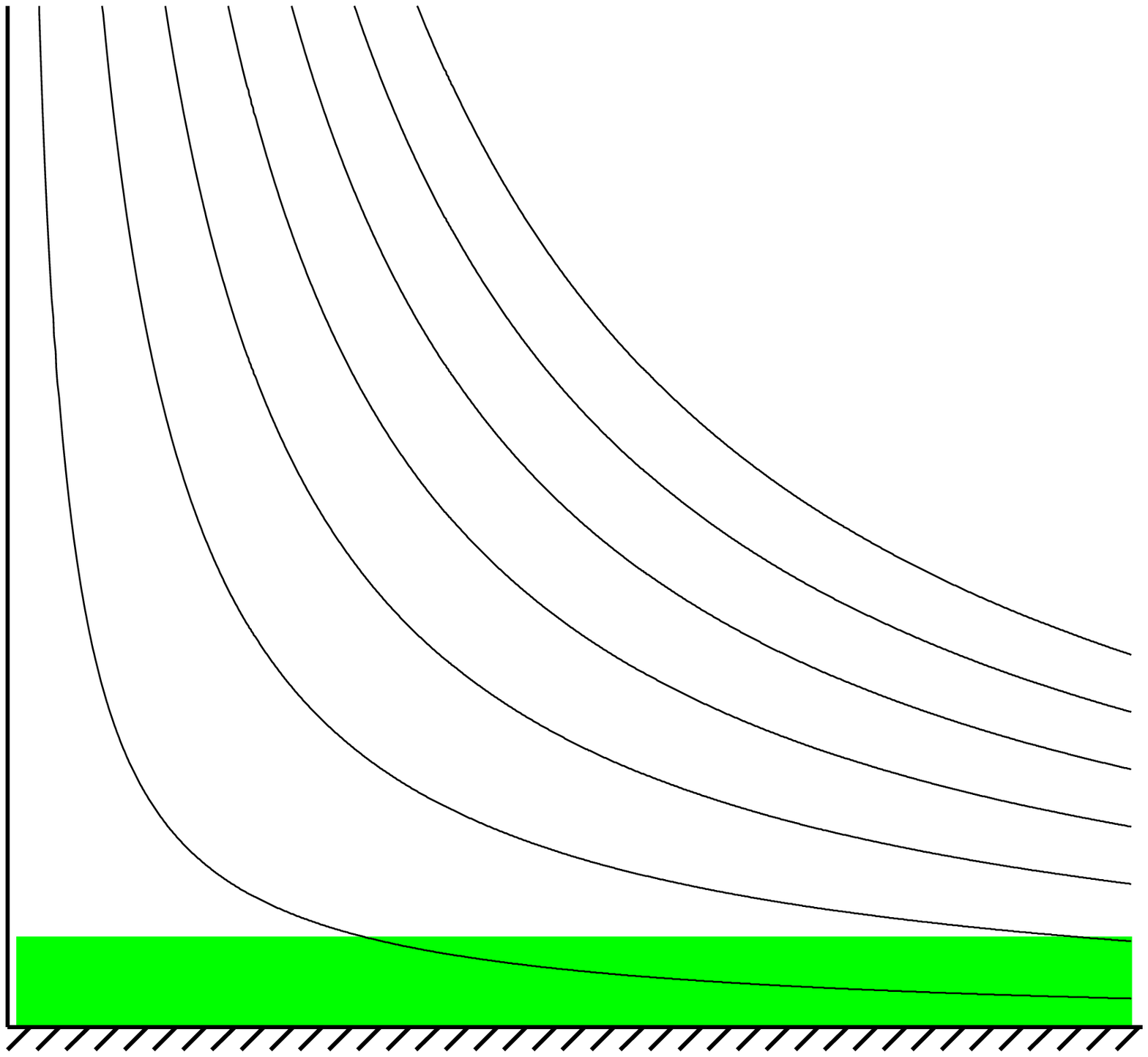}
\includegraphics[width=0.66\columnwidth,angle=90]{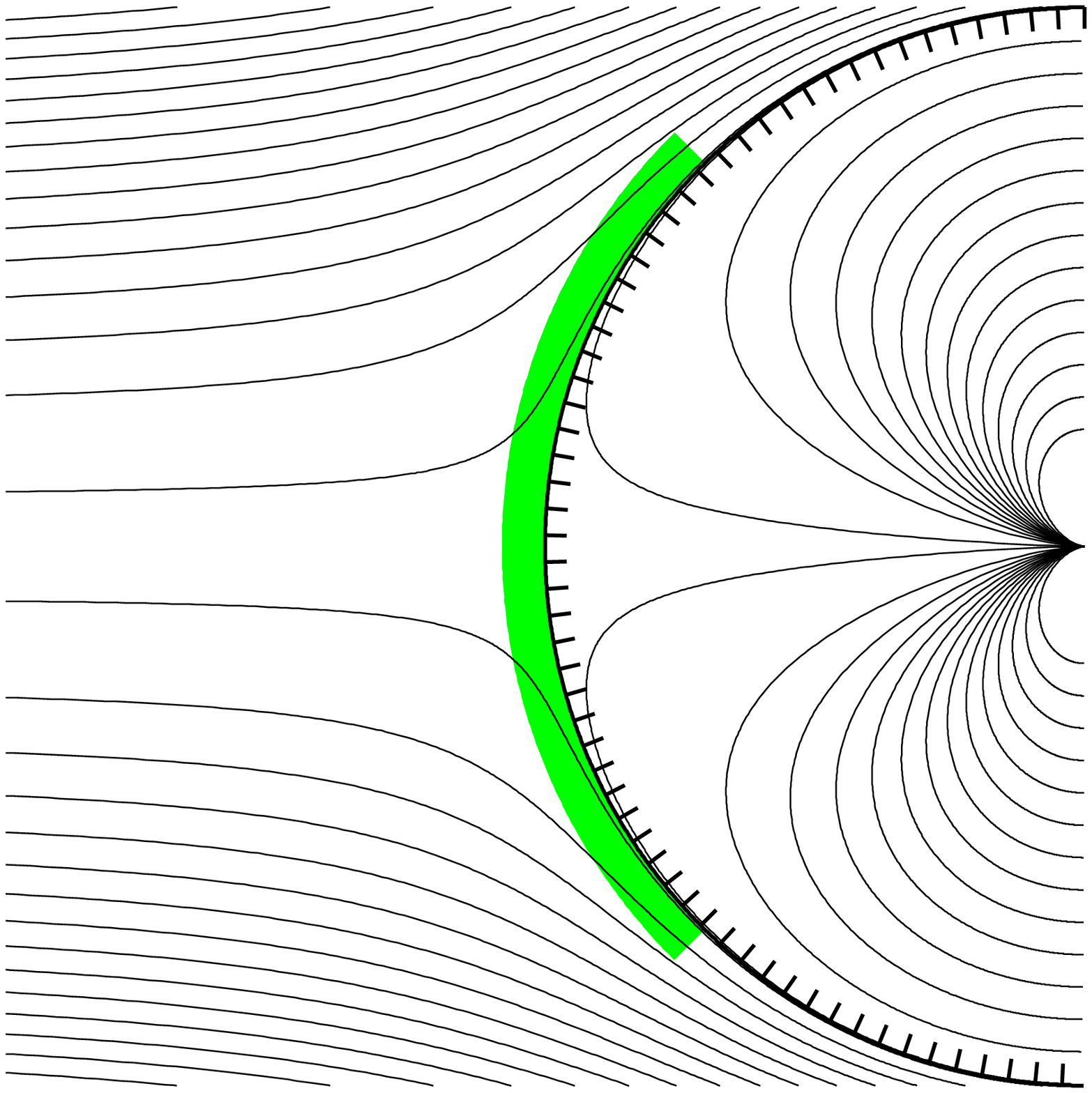}
\caption{Examples of simple potential flows: uniform flow past a plate
(left), flow into a 90 degree corner (middle) and flow past a sphere
(right). Streamlines are shown as thin solid lines. The shaded areas
schematically indicate the width of the layer formed by diffusion.
\label{fig:geo}
}
\end{figure*}
Churazov \& Inogamov (2004) noted that the behaviour of a conducting
layer in cold fronts should be similar to the behaviour of a viscous
layer near a plate or near the surface of a blunt body (see
e.g. Batchelor, 1967). When the fluid is advected along the surface,
the thickness of the layer grows in proportion to the square root of
the advection time. Near the stagnation point, the velocity of the flow
increases linearly with the distance from the stagnation point and the
characteristic advection time is approximately constant. Therefore the
thickness of the layer can also be approximately constant. Below we
provide a more rigorous justification of this picture.

Let us consider the simple case of diffusion of a passive scalar
$\psi$ in a potential flow of an incompressible fluid. The diffusion
coefficient $D$ is assumed to be constant\footnote{We use the notation
$D$ in this section for constant diffusion coefficient to distinguish
it from the temperature dependent heat conductivity $k$.} and the
velocity field is known and constant with time. The diffusion equation
\begin{eqnarray}
\frac{\partial \psi}{\partial t} + \vec{\nabla} \cdot ({\vec v} \psi) - D \Delta \psi=0
\label{eq:scalar0}
\end{eqnarray}
is supplemented by static boundary conditions at the surface of the
body and at large distance from the body.  For a steady state solution
($\frac{\partial \psi}{\partial t}=0$) and for an incompressible
fluid ($\vec{\nabla} \cdot \vec{v} = 0$) the above equation reduces to
\begin{eqnarray}
{\vec v} \cdot \vec{\nabla} \psi - D \Delta \psi=0.
\label{eq:scalar1}
\end{eqnarray}

In the simplest case of a uniform flow along the ``heated'' plate
(Fig.~\ref{fig:geo} left), ${v_x}=u={\rm const}$ and ${v}_y=0$. At
sufficiently large distance from the leading edge of the plate, the
derivative $\frac{\partial^2 \psi}{\partial^2 x}$ can be neglected and
equation (\ref{eq:scalar1}) can be written as
\begin{eqnarray}
u\frac{\partial \psi}{\partial x} -D \frac{\partial^2 \psi}{\partial y^2}=0.
\label{eq:e_flat}
\end{eqnarray}
An obvious solution in the form $\psi=f(y/\sqrt{x})$ is given by
\begin{eqnarray}
\psi=(\psi_1-\psi_2)~{\rm Erf}\left( \sqrt{\frac{u}{2 Dx}}y\right)+\psi_2,
\label{eq:s_flat}
\end{eqnarray}
where $\psi_2$ and $\psi_1$ are the values of the scalar at the plate
and at infinity, respectively. The width of the interface is therefore
$\Delta y=\sqrt{\frac{2 D}{u}x}$ and it increases with the distance
$x$ from the leading edge of the plate as $\sqrt{x}$. Since it takes a
time $t=x/u$ for the gas to flow from the edge of the plate to a given
position $x$, the width of the diffusive layer is simply $\sim
\sqrt{Dt}=\sqrt{Dx/u}$.

Consider now a potential flow into a 90 degrees corner
(Fig.~\ref{fig:geo} middle), governed by the velocity potential
$\phi=Ar^2\cos{2\theta}$ (see e.g. Lamb 1932, for various examples
of potential flows).  Here $r$ is the distance from the corner and
$\theta$ is angle from the horizontal axis. In this case the
velocity components are ${v}_x=2Ax$ and ${v}_y=-2Ay$. An obvious
solution to equation~(\ref{eq:scalar1}) is then
\begin{eqnarray}
\psi=(\psi_1-\psi_2)~{\rm Erf}\left( \sqrt{\frac{A}{D}}y\right)+\psi_2,
\label{eq:s_corner}
\end{eqnarray}
with the width $y=\sqrt{\frac{D}{A}}$ of the interface being
independent of $x$.  The reason for this behaviour is clear: the
acceleration of the (incompressible) fluid along the interface causes
a contraction of the fluid elements perpendicular to the direction of
the acceleration. While diffusion is trying to make the interface
broader, the motion of the fluid towards the interface compensates for
the broadening of the interface, and a steady state is reached
(Fig.~\ref{fig:geo} middle).

The potential flow past a cylinder or sphere behaves qualitatively
similar (Fig.~\ref{fig:geo} right). Indeed, in the vicinity of the
stagnation point (for $\theta \ll 1$), the radial and tangential
components can be written as (flow is from the right to the left,
angle is counted clockwise from the $-x$ direction):
\begin{eqnarray}
&&{ v}_r =-U\left(1-\frac{R^2}{r^2}\right)\cos{\theta}\approx
-2U\frac{\eta}{R} \nonumber \\
&&{v}_\theta = U\left(1+\frac{R^2}{r^2}\right)\sin{\theta}\approx
2U\frac{\zeta}{R},
\end{eqnarray}
for a cylinder and
\begin{eqnarray}
&&{ v}_r=-U\left(1-\frac{R^3}{r^3}\right)\cos{\theta}\approx
-3U\frac{\eta}{R} \nonumber \\ &&{v}_\theta=U\left(1+\frac{R^3}{2r^3}\right)\sin{\theta}\approx
\frac{3}{2}U\frac{\zeta}{R}
\end{eqnarray}
for a sphere. Here $U$ is the velocity at infinity, $R$
is the radius of the cylinder or sphere, $\eta=r-R$ and $\zeta=R\sin{\theta}$.

In the same approximation as for the cases discussed above (where the
spatial derivative of $\psi$ along $\zeta$ is neglected) the diffusion
equation reduces to
\begin{eqnarray}
-v_r \frac{\partial \psi}{\partial r} - D \frac{\partial^2
\psi}{\partial r^2}=0,
\label{eq:e_sphere}
\end{eqnarray}
and the width of the interface over the radius is set by the diffusion
coefficient $D$ and the coefficient $C$ in the relation
$v_r=-C\eta$, yielding
\begin{eqnarray}
\Delta r\approx \sqrt{\frac{2 D}{C}}=\left\{\begin{array}{ll}
\sqrt{D\frac{R}{U}}& {\rm cylinder} \\
&\\
\sqrt{\frac{2}{3}D\frac{R}{U}}& {\rm sphere.}
\end{array}
\right.
\label{eq:e_sphere2}
\end{eqnarray}
In this case the width of the interface is also constant along the
surface of the cylinder or sphere (Fig.~\ref{fig:geo} right).

One can also consider a closer analogue of a flow past a spherical cloud by
extending the solution for a potential flow into the inner part of the
cylinder or sphere, as illustrated in Fig.~\ref{fig:geo}. In this model there
is a circulation flow of gas inside the cloud, and the tangential component of
the velocity is continuous across the boundary while the normal component is
zero at the boundary. We can further allow for different densities $\rho_1$
and $\rho_2$ outside and inside of the boundary if all velocities inside are
scaled by a factor $\sqrt{\rho_1/\rho_2}$. The resulting configuration can be
considered as an idealized (and unstable) analogue of a hot flow past a colder
cloud in the absence of gravity (see also Heinz et al., 2003). Allowing different
diffusion coefficients $D_1$ and $D_2$ in the flow outside and inside the
boundary, and requiring the solution $\psi$ and its spatial derivative to be
continuous across the interface, yields the following solution in the vicinity
of the stagnation point:
\begin{eqnarray}
\psi&=&(\psi_1-\psi_m)~{\rm Erf}\left(
\sqrt{\frac{C_1}{2D_1}}(r-r_0)\right)+\psi_m \;\; {\rm outside,}
\nonumber \\ \psi&=&(\psi_m-\psi_2)~{\rm Erf}\left(
\sqrt{\frac{C_2}{2D_2}}(r-r_0)\right)+\psi_m, \;\;{\rm inside}
\nonumber \\
\psi_m&=&\frac{\psi_1+\psi_2\frac{D_2}{D_1}\frac{C_1}{C_2}}{1+\frac{D_2}{D_1}\frac{C_1}{C_2}}.\nonumber
\end{eqnarray}
Here $r_0$ is the radius of the boundary, $\psi_1$ and $\psi_2$ are the
values far from the interface, $C_2=C_1\sqrt{\rho_1/\rho_2}$, and $C_1=2U/R$
for a cylinder or $C_1=3U/R$ for a sphere, respectively. The width of the
interface is again constant along the boundary.

The same answer is obviously valid for any idealized flow of this
type: near the stagnation point the width of the ``heated'' layer
does not change along the surface of the body. Real cold fronts
are of course much more complicated structures. However, the
acceleration of the flow along the interface and the simultaneous
contraction in the perpendicular direction are generically present
also here. It can therefore be expected that the width of the
interface will be similarly constant in real cold fronts. A simple
extension of the above toy model can be obtained by allowing for
gas compressibility and a temperature dependent conductivity, i.e.
by considering the full system of equations
(\ref{eq:hydro1})-(\ref{eq:hydro}) with conductivity according to
eq.~(\ref{eq:k0}). An expansion of heated layers and simultaneous
contraction of cooled layers on the other side of the interface
will certainly modify the flow, but for the transonic flows of
interest here we might expect that the results obtained for a toy
model will still be approximately valid. In the next section we
verify this prediction using numerical simulations.

\section{Numerical simulations}
\label{sec:num}
For our numerical experiments, we used the TreeSPH code {\small GADGET-2}
(Springel, 2005) combined with the implementation of thermal conduction
discussed by Jubelgas, Springel \& Dolag (2004), which accounts both for the
saturated and unsaturated regimes of the heat flux.

The simulations were intended to illustrate a simple toy model,
described in section \ref{sec:toy}, rather than to provide a
realistic description of the observed cold fronts. The specific
goal was to see the impact of the flow stretching near the
stagnation point on the width of the interface set by conduction.
With this in mind we intentionally restricted ourselves to a 2D
geometry and an unmagnetized plasma. For a 3D calculation of
magnetized clouds see Asai et al.~(2007).  The self gravity of gas
particles was also neglected in our idealized simulations and all
gas motions were happening in a static gravitational potential.
Given that the typical gas mass fraction in clusters is of order
10-15 per cent, the self gravity of gas particles is likely to be
a second order effect. A more significant simplification is the
assumption of a static potential, since at least some of the cold
fronts are caused by cluster mergers where strong changes of the
potential are possible. Formation of cold fronts in the
appropriate cosmological conditions was considered by e.g. Bialek
et al. (2002), Nagai \& Kravtsov (2003), Mathis et al. (2005), see
also Tittley \& Henriksen (2005) and Ascasibar \& Markevitch
(2006).  Our illustrative 2D simulations, described below, can be
viewed as a ``minimal'' configuration which allows us to see the
effect of flow stretching and to extend the toy model to the case
of a compressible gas and a temperature dependent diffusion
coefficient.

\subsection{Initial conditions}
Our 2D simulations of cold fronts in clusters were carried out in a 8x4 Mpc
periodic box. We represented the cluster with a static King gravitational
potential of the form
\begin{equation}
\phi=-9\sigma^2 \frac{\ln{\left[
      x+\sqrt{1+x^2}\right]}}{x},
\end{equation}
with $\sigma = 810\,{\rm km\,s^{-1}}$,
$x=r/r_c$ and $r_c=300$ kpc. The initial temperature and density distributions
were set to
\begin{eqnarray}
(T_e, \rho)=\left\{\begin{array}{ll}
(T_1, \frac{\rho_1}{ (1+x^2)^\frac{3}{2} })& x<x_{\rm out} \\
& \\
(T_2, \frac{\rho_2}{(1+x^2)^{\frac{3}{2}\frac{T_1}{T_2}}})& x>x_{\rm out}
\end{array}
 \right.
\end{eqnarray}
where
\begin{equation}
\rho_2=\frac{\rho_1}{
(1+x_{\rm out}^2)^\frac{3}{2}}\frac{T_1}{T_2}\left[
(1+x_{\rm out}^2)^\frac{3}{2}\frac{T_1}{T_2} \right],
\end{equation}
 and $kT_1=\mu
m_p\sigma^2\approx 4$ keV. Thus the temperature and density make a
jump at $x_{\rm out}$, while the pressure is continuous. In our runs,
$T_2=8$ keV, $x_{\rm out}=1$, $\mu=0.61$, and
$\rho_1=6.6\times10^{-26}~{\rm g~cm^{-3}}$.  The gas velocity was
set to zero for $x<x_{\rm out}$ and to $u=2000~{\rm km~s^{-1}}$ for
$x>x_{\rm out}$. The corresponding Mach number relative to the hot 8
keV gas is $\sim$1.3 (neglecting further acceleration of the flow
in the cluster potential).

$10^6$ gas particles of equal mass were distributed over the
computational volume as a Poissonian sample of the initial density
distribution described above. The gas temperature for each
particle was set to $T_1$ or $T_2$ depending on the position of
the particle. The Poissonian noise introduced by this procedure
leads to small-scale (and small amplitude) pressure/entropy
perturbations in the initial conditions. Shortly after the
beginning of the simulations, the over-dense regions expand and
create a pattern of ripples in the temperature distribution (see
Fig.~\ref{fig:snap}, top panel). We stress that the presence of
these ripples is a direct consequence of the choice of initial
particle positions and it does not mean that the number of
particles is insufficient to properly resolve the cold front.  A
comparison of three runs with $3~10^5$, $10^6$ and $3~10^6$
particles with the same initial condictions and the conduction
suppression coefficient $f=0.01$ yielded practically
undistinguishable results in terms of the cold front structure. As
these ripples do not affect the overall structure of the flow no
attempt was made to correct the initial conditions for this
effect.  These ripples are also a useful visual indicator of the
impact of thermal conduction on the small-scale temperature
structures in the flow (see Fig.~\ref{fig:snap}).

Our choice of initial conditions has been motivated by the cold front in the
cluster Abell 3667 (Vikhlinin et al., 2001), but we did not try to accurately
reproduce all the observed properties of this cluster. In particular, the
location and the strength of the shock in our model need not be the same as
in A3667.  Nevertheless, the most important feature of A3667 -- a cool gas
cloud inside a hotter and less dense flow -- is present in our simulations,
allowing us to study the impact of thermal conduction on the interface between
the cloud and the flow. To this end we carried out four runs where the
coefficient of the thermal conduction efficiency was set to $f=0$, $0.01$,
$0.1$, and $0.5$, respectively.

\subsection{Results}
\label{sec:res}

\begin{figure}
\centering
\begin{tabular}{c}
\begin{minipage}{0.42\textwidth}
\centering
\includegraphics[width=\textwidth]{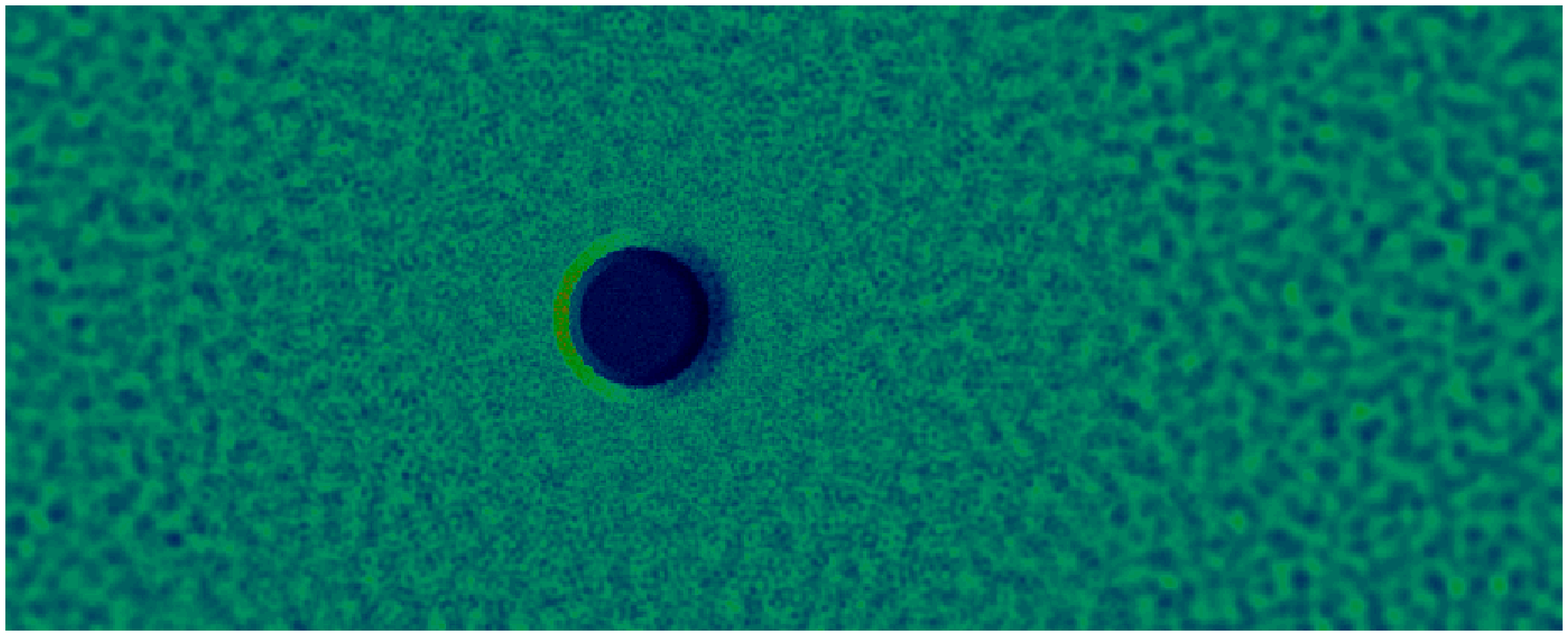}
\end{minipage}
\\
\begin{minipage}{0.42\textwidth}
\centering
\includegraphics[width=\textwidth]{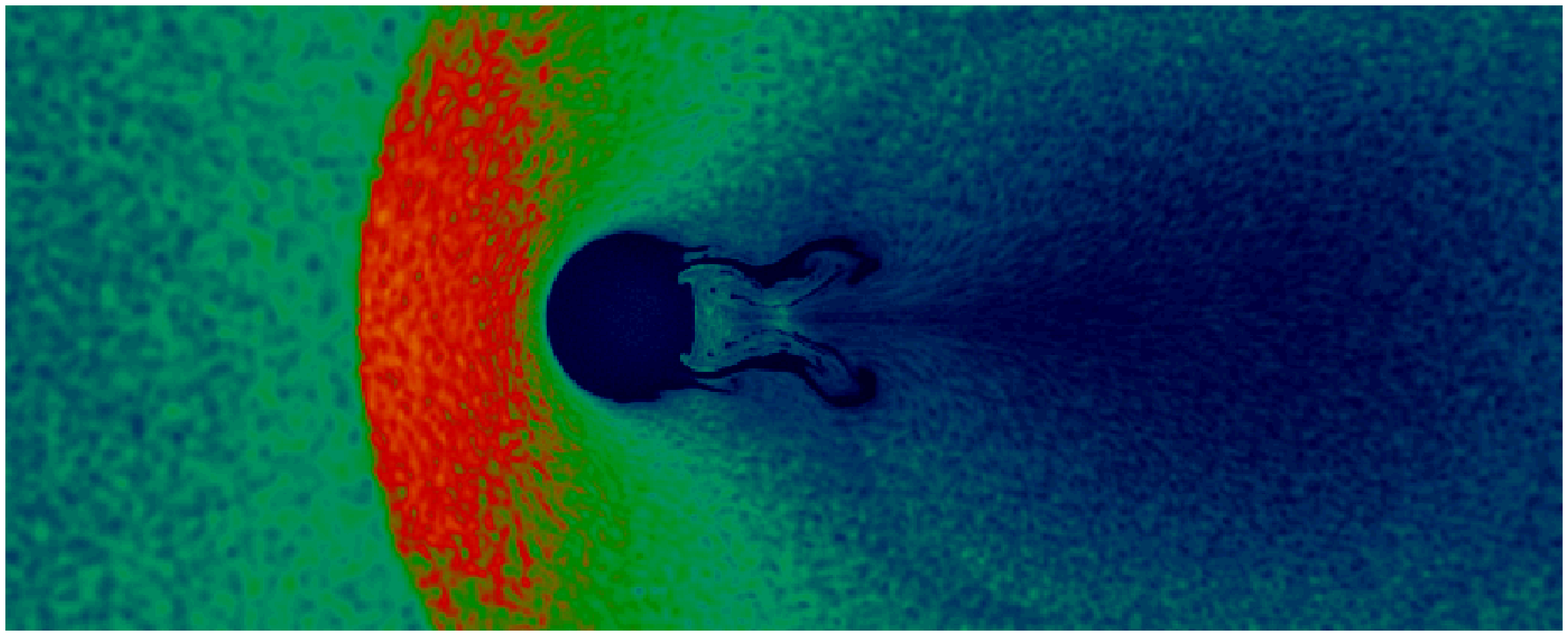}
\end{minipage}
\\
\begin{minipage}{0.42\textwidth}
\centering
\includegraphics[width=\textwidth]{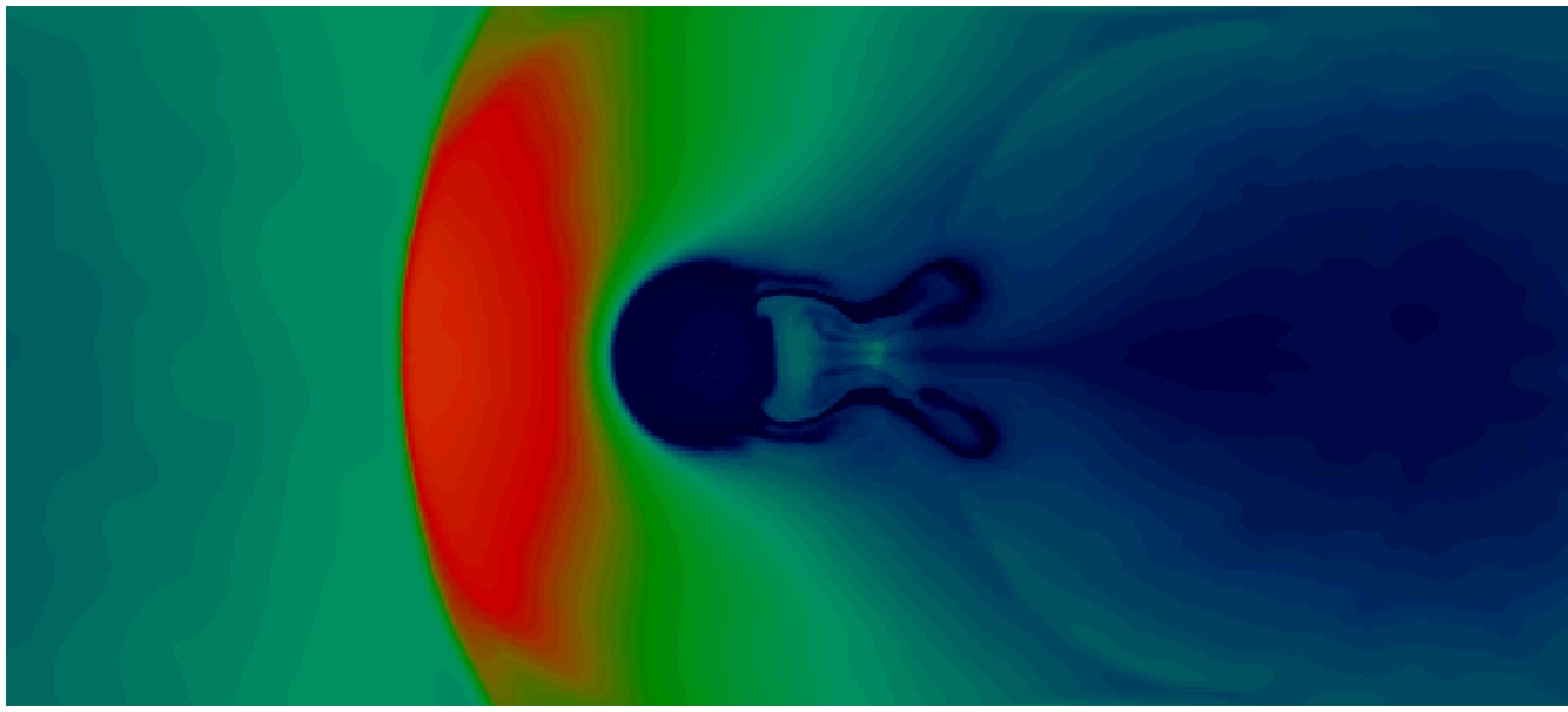}
\end{minipage}
\\
\begin{minipage}{0.42\textwidth}
\centering
\includegraphics[width=\textwidth]{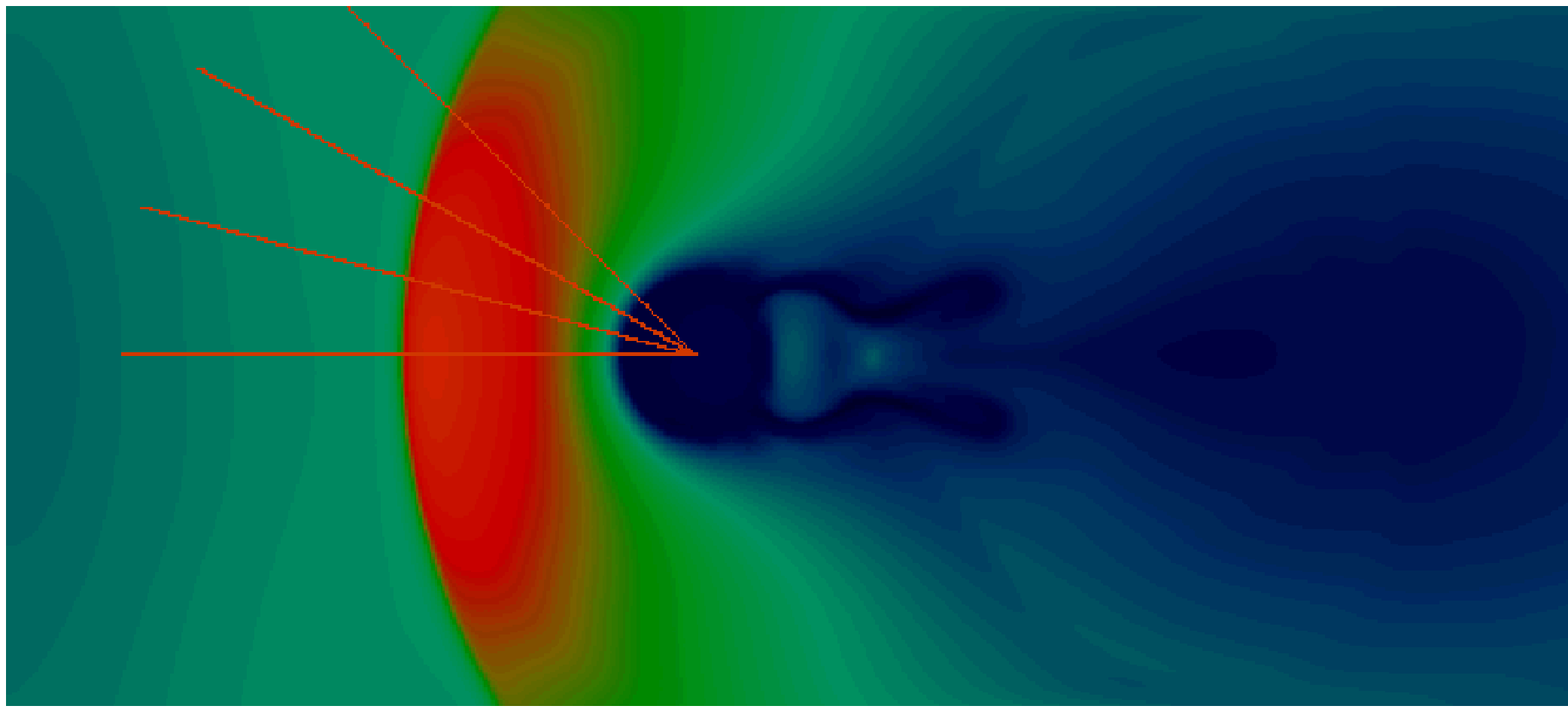}
\end{minipage}
\\
\begin{minipage}{0.42\textwidth}
\centering
\includegraphics[width=\textwidth]{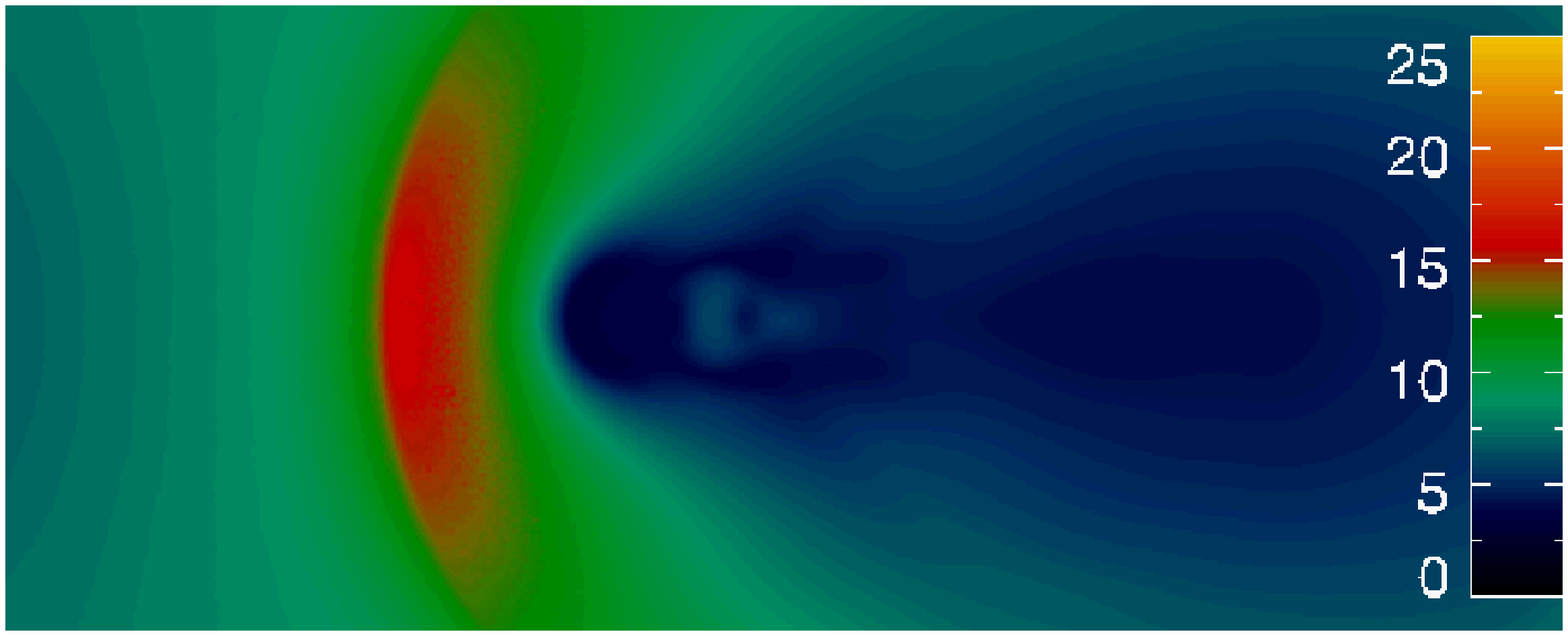}
\end{minipage}
\end{tabular}
\caption{Snapshots of the temperature distribution for a 2D flow of
hot (8 keV) gas past a cold (4 keV) cloud. The image sizes are 6.7 by
2.7 Mpc. The top panel corresponds to a moment shortly after the start
of the simulations. In the other panels, the temperature distributions
at $t=1.3\,{\rm Gyr}$ after the beginning of the simulations are shown
as a function of the strength of thermal conduction.  The conduction
suppression coefficient for these panels is $f=0$, $0.01$, $0.1$, and
$0.5$, respectively. Thus the second panel shows the run without
conduction, while the bottom panel corresponds to a conductivity equal
to half the Spitzer-Braginskii value. The temperature structure of the
interface plotted in subsequent figures was measured along the red
lines shown in one of the panels.}
\label{fig:snap}
\end{figure}

In Fig.~\ref{fig:snap}, we show the temperature distributions for all
four runs 1.3 Gyr after the start of the simulations. Immediately
after the beginning of the simulations a shock starts to propagate
upstream through the hot flow, forming a clearly visible bow shock.
Because of the acceleration in the cluster potential, the Mach number
of the shock is $\sim$1.7 (rather than 1.3) and the temperature behind
the shock is also rather high (10-15 keV).

\begin{figure*}
\includegraphics[width=1.9\columnwidth]{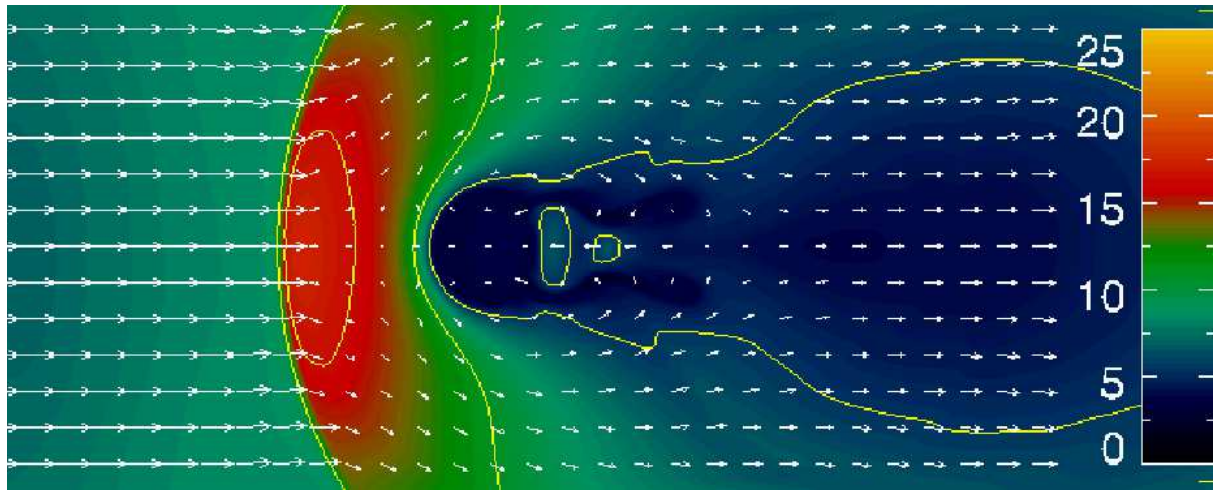}
\caption{Snapshot of the temperature distribution at $t=1.3\,{\rm
Gyr}$ after the start of a simulation with conductivity coefficient
$f=0.1$, with the velocity field superposed.}
\label{fig:jv}
\end{figure*}

The cool cloud is first pushed back by the ram pressure of the gas and then
(slowly) oscillates near an equilibrium position. At 1.3 Gyr, there are still
some residual motions clearly associated with the specific initial conditions,
but these motions are quite gradual. This can also be seen in the gas velocity
field, which is plotted in Fig.~\ref{fig:jv}. It shows a clearly visible
velocity jump at the shock front, and inside and around the cloud, circular
motions are present, broadly resembling the velocity field shown in
Fig.~\ref{fig:geo}. Such circular motions inside the cloud lead to a transport
of low entropy gas from the centre of the cloud towards the stagnation point
(Heinz et al., 2003). As a result of adiabatic expansion of the transported
gas its temperature drops to $\sim$ 3 keV, below the initial value of 4 keV.

\begin{figure}
\includegraphics[width=\columnwidth]{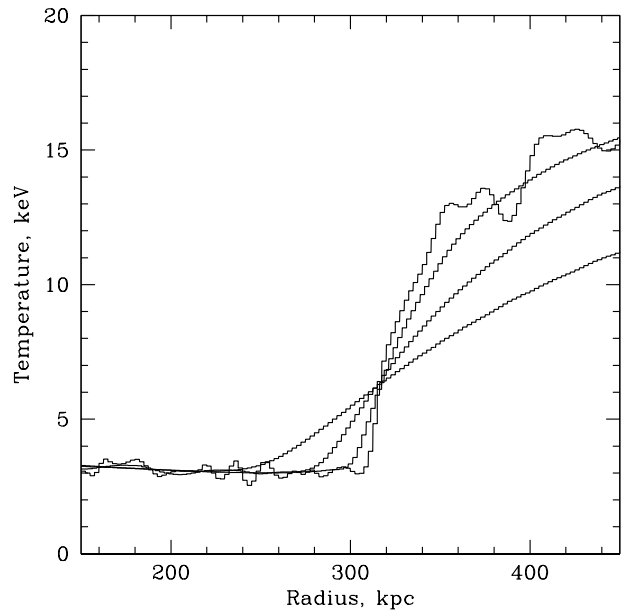}
\caption{Temperature profiles along the symmetry axis of the cloud at
$t=1.3\,{\rm Gyr}$ after the start of the simulations, for different
conduction suppression coefficients equal to $f=0$, $0.01$, $0.1$, and
$0.5$. The distance is measured from the centre of the cold cloud.}
\label{fig:profk}
\end{figure}

Since after 1.3 Gyrs much of the relaxation from the initial state
already took place, we compare the runs with different conductivity at
this time.  The effect of increasing the efficiency of thermal
conduction is clearly visible in the snapshots shown in
Fig.~\ref{fig:snap}.  First of all, small scale temperature variations
present in the initial conditions are smoothed out in all runs where
thermal conduction is present. Secondly, with the increase of $f$ the
interface separating the cloud and the hot flow becomes less and less
sharp. This is seen more explicitly in the temperature profiles across
the interface (along the symmetry axis of the cloud), which are shown
in Fig.~\ref{fig:profk}. In this figure (and in the subsequent
figures), the distance (plotted along the abscissa axis) is measured
from the approximate centre of the cloud. Since the cloud is not
perfectly spherical, its position varies slowly with time. As its
centre is hence not accurately known, all profiles shown in
Fig.~\ref{fig:profk} were shifted along the abscissa axis to have the
temperature value 6 keV gas at the same position.

The sharpest profile corresponds of course to the run without conduction, and
in this case some small-scale fluctuations left over from the initial
conditions can still be seen in the profile. This run also sets a useful
benchmark for comparison with the other simulations, for example, it indicates
the numerical resolution available for representing the interface. We see
that for values of $f$ larger than 0.01 the impact of thermal conduction on
the width of the interface can be well resolved with our numerical setup.  For
runs with $f=0.01$, $0.1$ and $0.5$, the small-scale fluctuations in the
temperature distribution are absent and the effective thickness of the
interface gradually increases.

\begin{figure}
\includegraphics[width=\columnwidth]{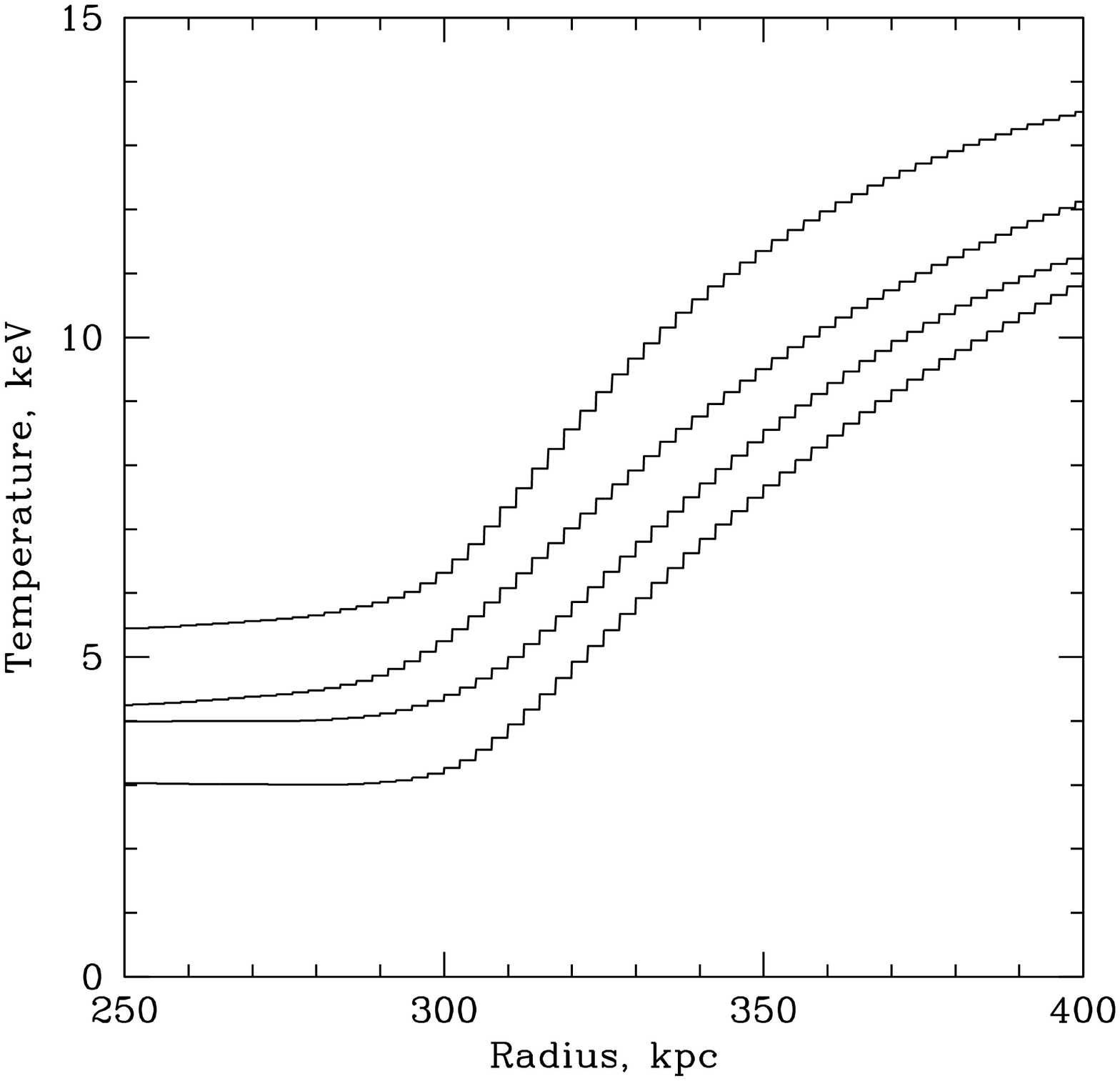}
\caption{Temperature profiles along the symmetry axis of the cloud
after $t=0.2$, $0.4$, $0.8$ and $1.3\,{\rm Gyr}$ since the start of
the simulations. The conduction suppression coefficient was set to
$f=0.1$. The distance is measured from the centre of the cloud. In
order to compensate for the gradual changes in the interface shape,
each profile was shifted along the X-axis so that the rising part of
the temperature profile has the same abscissa for all curves.}
\label{fig:proftime}
\end{figure}

We can now verify our simple predictions based on the toy model of
diffusion of a passive scalar in a potential flow. We first consider
our finding that after an initial settling time of order $R/U$ the
interface evolves to a quasi-steady state.  This is indeed seen in
Fig.~\ref{fig:proftime}, where the temperature profiles along the
symmetry axis are shown for $t=0.2$, $0.4$, $0.8$, and $1.3\,{\rm
Gyr}$ since the beginning of the simulations. While there is clear
evolution of the profile (e.g. in terms of the maximal or minimal
temperatures) the shape of the interface is very similar at all times.

Another expectation is that the thickness of the interface is the same
along the interface (as long as the distance from the stagnation point
is much less the the cloud curvature radius). Indeed, the profiles
measured at different distances from the stagnation point
(Fig.~\ref{fig:proft}) look very similar. In this figure, the profiles
were calculated along directions making different angle with respect
to the symmetry axis of the cloud (red lines in
Fig.~\ref{fig:snap}). This means that when deriving an effective
conduction coefficient from the observed cold fronts one can use the
profile averaged over a large part of the interface, rather than being
constrained to small sectors of the front.

\begin{figure}
\includegraphics[width=\columnwidth]{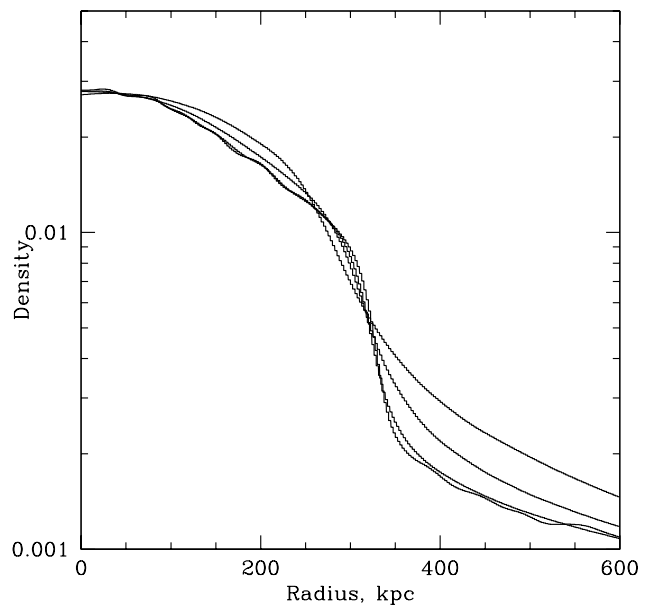}
\caption{Gas density profiles along the symmetry axis of the cloud at
$t=1.3\,{\rm Gyr}$ after the start of the simulations, for different
conduction suppression coefficients equal to $f=0$, $0.01$, $0.1$, and
$0.5$. The distance is measured from the centre of the cold cloud.}
\label{fig:profd}
\end{figure}

Thus the results of the numerical simulations are broadly consistent
with the expectations derived earlier: once the front is formed, it
has a width constant in time and constant along the interface.

\section{Simple estimates of the interface width}

We are now looking for a simple 1D problem for a compressible fluid
which has a solution that can provide a qualitative approximation of
the cold front structure. The analogy with the problems considered in
Section~\ref{sec:toy} suggests that the width of the interface should
scale as $\sqrt{D t_s}$, where $D$ is the effective diffusion
coefficient (thermal conductivity) and $t_s$ is the effective time
scale.  For the flows in Section \ref{sec:toy}, a reasonable choice
was $t_s\sim R/U$, where $R$ is the curvature radius of the interface
and $U$ is the velocity of the flow at infinity. Of course, this
result was derived for a potential flow, and for more realistic cases
it might be more correct to recast $t_s$ in the form
$t_s=\left(\frac{{\rm d} v_r}{{\rm d} r} \right)^{-1}$, where $v_r$ is
the velocity component perpendicular to the interface. Indeed, from
Eqs.~(\ref{eq:e_sphere}) and (\ref{eq:e_sphere2}) it is clear that
that this quantity (i.e. the gradient of the radial velocity
component) enters the expression of the interface width. We can hence
try to obtain an approximate solution for the front structure by
considering a 1D time dependent diffusion equation\footnote{An
alternative approach is to incorporate the velocity field of the base
flow into the energy conservation equation and to look for a steady
state solution.}, starting from a Heaviside step function for the
initial temperature distribution and taking the solution at time
$t_s$. The hope is that for this choice of $t_s$ the most basic
properties of the interface structure will be captured.

We also assume that all velocities in the vicinity of the interface
are  small compared to the sound speed, all quadratic
terms in $v$ can be neglected and that the pressure is approximately
constant across the interface. Then the gas density is $\rho=\rho_0
\frac{T_0}{T}$, where $\rho_0 T_0=P_0$ is fixed by the initial
pressure.  In this approximation, the heat diffusion equation reduces
to
\begin{eqnarray}
  \frac{\partial T}{\partial t}=\lambda\frac{\partial }{\partial x} k   \frac{\partial T}{\partial x} - \lambda \frac{k}{T} \left( \frac{\partial T}{\partial x} \right)^2,
\label{eq:lambda}
\end{eqnarray}
where $\lambda=\frac{\gamma-1}{\gamma}\frac{\mu m_p}{\rho k_B}$. This
equation is very similar to the standard diffusion equation in solids,
except for the second term in the r.h.s. which accounts for gradual
expansion of the heated gas, and for the contraction of the cooled gas
in order to maintain constant pressure across the interface.

\begin{figure}
\includegraphics[width=\columnwidth]{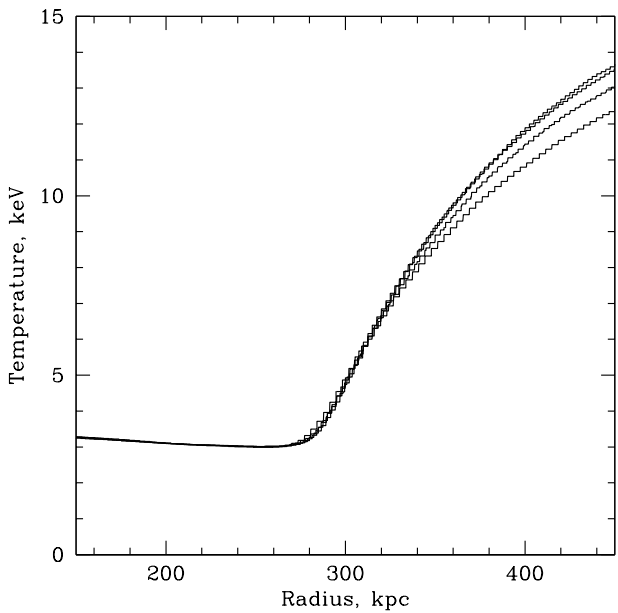}
\caption{Temperature profiles at 1.3 Gyr for $f=0.1$ along directions
making an angle of 0, 15, 30, or 45 degrees with respect to the
symmetry axis of the cloud (see also Fig.~\ref{fig:snap}).}
\label{fig:proft}
\end{figure}

Equation~(\ref{eq:lambda}) can be readily integrated. We set the
initial values of the temperature to 3 and 15 keV on the two sides of
the interface, respectively, and the electron density to
$10^{-2}\,{\rm cm}^{-3}$ on the cool side to approximately reproduce
the properties of the simulated front, as shown in
Fig.~\ref{fig:proft}.

\begin{figure}
\includegraphics[width=\columnwidth]{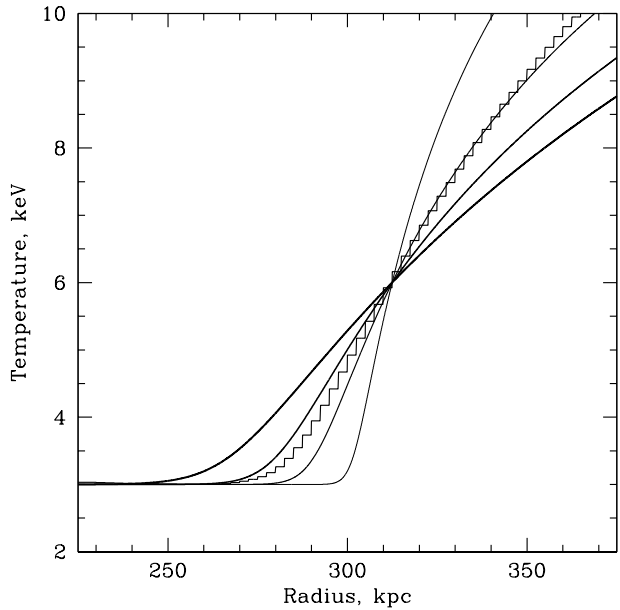}
\caption{Comparison of the temperature profile along the symmetry axis
of the cloud (for $f=0.1$) obtained in numerical simulations
(histogram) and obtained from equation~(\ref{eq:lambda}) for
$t_s=0.1$, $0.4$, $0.9$, and $2.0\,{\rm Gyr}$.}
\label{fig:profan}
\end{figure}

In Figure~\ref{fig:profan}, we plot the solution of the equation at
times $t=0.1$, $0.4$, $0.9$, and $2.0\,{\rm Gyr}$ for $f=0.1$ together
with the temperature profile obtained in the SPH simulations. The
results of the numerical simulations best correspond to the solution
of Eqn.~(\ref{eq:lambda}) for $t\sim 0.5-0.7\,{\rm Gyr}$. For
comparison, the ratio of the cloud radius to the flow velocity at
infinity is $R/U\sim 1.5\times10^8$ yr, while $t_s$ evaluated from the
velocity profile obtained in the simulations is $t_s=\left(\frac{{\rm
d} v_r}{{\rm d} r} \right)^{-1}\approx 5\times10^8\,{\rm yr}$. The
difference in the estimated width of the front based on the $R/U$
ratio compared with a more detailed treatment of the velocity field is
of order $2$. This discrepancy (for our numerical setup) is largely
caused by i) the drop of the velocity at the shock and ii) differences
between the velocity field obtained in the simulations and that in
potential flows of incompressible fluids, as considered in Section
\ref{sec:toy}.  Of course, Eqn.(\ref{eq:lambda}) by itself is only a
crude approximation of the problem. Nevertheless, even our simplest
estimate predicts the width of the interface within a factor of 2 of
the value derived from direct numerical simulations.

Note also that because of the expansion of the heated gas, the actual
contact discontinuity does not necessarily exactly coincide with the
``visible'' boundary of the cool cloud. Indeed, colder gas of the
cloud expands while the hotter ambient gas contracts, shifting the
discontinuity away from the cloud centre. At the same time, the
sharpest edge will be observed in places where the temperature
gradient is large. If the gas on the two sides of the contact
discontinuity has different abundances of heavy elements then the true
position of the contact discontinuity can be determined from the
abundance gradient. This exercise however requires data of very high
quality.

We thus find that an order of magnitude estimate of the width of a
spherical cold front can be written as
\begin{eqnarray}
  \Delta r\approx \delta \sqrt{\frac{2}{3}D\frac{R}{U}}=\delta \sqrt{\frac{2}{3}\frac{\gamma-1}{\gamma}\frac{\mu m_p}{\rho}\frac{f k_0}{k_B}\frac{R}{U}},
\end{eqnarray}
where the factor $\delta$ accounts for all departures introduced by
the approximations involved in our simplest model considered in
Section~\ref{sec:toy}. Based on our numerical simulations we have
$\delta\sim 0.5$. Plugging in fiducial values for the A3667 cluster
(Vikhlinin et al., 2001) one gets:
\begin{eqnarray}
  \Delta r\approx 40 \left(\frac{\delta}{0.5}\right) f^{0.5} T_5^{5/4}
  R_{300}^{0.5} U_{1400}^{-0.5}
  N_{0.002}^{-0.5}~\;{\rm kpc},
\end{eqnarray}
where $T_5=\frac{T}{5\,{\rm keV}}$, $R_{300}=\frac{R}{300\,{\rm kpc}}$,
$U_{1400}=\frac{U}{1400\,{\rm km\,s^{-1}}}$, and $N_{0.002} = \frac{n_e}{2\times
10^{-3}\,{\rm cm}^{-3}}$.

This value is a factor of $\sim 8$ larger than the upper limit for the
interface width derived from Chandra observations. If this discrepancy
is solely caused by a suppression of thermal conduction, then the
factor $f$ has to be less than $1.5\times10^{-2}$.  Note that the mean
free path is $\lambda\sim 4~ T_5^2 N_{0.002}^{-1}\, {\rm kpc}$
(e.g. Sarazin, 1986), which is smaller than the interface width
evaluated for $f=1$. If the effective mean free path of electrons
scales linearly with $f=k/k_{0}$ (while the interface width $\Delta r
\propto \sqrt{f}$) then for all $f<1$ the width of the interface will
remain larger than the mean free path. Therefore our assumption of
unsaturated heat flux remains valid. For the observed cold front in
A3667 the effective mean free path of electrons should therefore be
factor $f=1.5\times10^{-2}$ smaller than in unmagnetized plasma.

However, we caution that magnetic fields are likely playing a role in
the structure of the interface as suggested by theoretical arguments
(e.g. Vikhlinin et al., 2001, Lyutikov 2006) and numerical simulations
(e.g. Asai 2004, 2005). Particularly important is the stretching of
the field lines along the interface, which can suppress heat
conduction across the front or even affect the hydrodynamical
stability of the interface.  Note that the heat conductivity depends
strongly on the topology of the magnetic field since the electron
Larmor radius is some 10 orders of magnitude smaller than the
characteristic length scales of the problem for typical magnetic field
strengths at the micro-Gauss level.

We note that the interface developing in our SPH simulations may also
be stabilized to some extent against small-scale fluid instabilities
by numerical effects. Across strong density discontinuities, SPH has
been found to produce spurious pressure forces that may suppress small
wavelength Kelvin-Helmholtz instabilities (Agertz et al. 2006).  This
effect is equivalent to a small surface tension and mimics the
stabilizing influence expected from an ordered magnetic field across
the front. However, better numerical resolution weakens this effect
and should allow ever smaller wavelengths to grow.

\section{Conclusions}

We have shown that in the presence of thermal conduction the width
of the interface separating hot gas flowing past a cooler gas
cloud (a ``cold front'' in clusters of galaxies) can be estimated
from the size $R$ of the cloud, the velocity $U$ of the gas and
the effective thermal conductivity. The structure of the interface
is established over a period of time $\sim R/U$, while the
subsequent evolution is much slower. Moreover, the width of the
interface is approximately constant along the front. We made an
illustrative 2D simulations of an unmagnetized plasma flow past a
colder cloud with gas densities and temperatures characteristic
for the observed cold fronts. While being very idealized, the
simualtions do show that the width remains approximately constant
when the gas compressiblity and the temperature dependence of the
conductivity is accounted for.

This implies that one can use much of the visible part of the
interface in order to assess the effective thermal conductivity of the
gas. For the cold front in Abell 3667, the estimated width of the
interface is $40 f^{0.5}~{\rm kpc}$, where $f$ is the conduction
suppression coefficient (relative to the Spitzer-Braginskii
value). This factor $f$ has to be smaller than $0.015$ in order to
reproduce the observed limits on the width of the interface.  This
result is consistent with previous suggestions that magnetic fields
play an important role in providing thermal isolation of the gases
separated by the cold front. The idealised description of the
interface presented here provides a useful method for estimating the
effective gas conductivity from observations of clusters of galaxies.

\section*{Acknowledgments}
EC is grateful to Nail Inogamov, Maxim Lyutikov and Maxim Markevitch
for useful discussions.

\label{lastpage}
\end{document}